\documentclass{emulateapj}

\newcommand{\mitrgb}{$M_I^{TRGB}$}
\newcommand{\msun}{$M_{\odot}$}
\newcommand{\itrgb}{$I_{TRGB}$}
\newcommand{\ebv}{E(B--V)}
\newcommand{\ai}{A$_I$}
\newcommand{\av}{A$_V$}

\newcommand{\zsol}{$Z_{\odot}$}
\slugcomment{Astrophysical Journal Letters, submitted, July 2008}

\shorttitle{The Distance of NGC 1569}
\shortauthors{Grocholski et al.}

\begin{document}


%\title{Deep HST/ACS Photometry of the Starburst Galaxy NGC 1569:\\ a 
%New Distance via the Tip of the Red Giant Branch\altaffilmark{1}}

\title{A New Hubble Space Telescope Distance to NGC 1569:\\
Starburst Properties and IC 342 Group Membership}


\author{Aaron J. Grocholski\altaffilmark{2}, Alessandra
Aloisi\altaffilmark{2,3}, Roeland P. van der Marel\altaffilmark{2},
Jennifer Mack\altaffilmark{2}, Francesca Annibali\altaffilmark{2}, Luca
Angeretti\altaffilmark{4}, Laura Greggio\altaffilmark{5}, Enrico V.
Held\altaffilmark{5}, Donatella Romano\altaffilmark{4}, Marco
Sirianni\altaffilmark{2,3}, Monica Tosi\altaffilmark{4}}


\altaffiltext{1}{Based on observations with the NASA/ESA {\it Hubble
Space Telescope}, obtained at the Space Telescope Science Institute,
which is operated by AURA for NASA under contract NAS 5-26555}

\altaffiltext{2}{Space Telescope Science Institute, 3700 San Martin Dr.,
Baltimore, MD 21218, USA; aarong, aloisi, marel, mack, 
annibali,sirianni@stsci.edu}

\altaffiltext{3}{On assignment from the Space Telescope Division of the
European Space Agency}

\altaffiltext{4}{Osservatorio Astronomico di Bologna, INAF, Via Ranzani
1, I-40127 Bologna, Italy; langeretti@gmail.com, donatella.romano, 
monica.tosi@oabo.inaf.it}

\altaffiltext{5}{Osservatorio Astronomico di Padova, INAF, vicolo 
dell'Osservatorio 5, 35122 Padova, Italy; laura.greggio, 
enrico.held@oapd.inaf.it}

\begin{abstract}
We present deep HST ACS/WFC photometry of the dwarf irregular galaxy NGC
1569, one of the closest and strongest nearby starburst galaxies. These
data allow us, for the first time, to unequivocally detect the tip of
the red giant branch and thereby determine the distance to NGC 1569.  We
find that this galaxy is 3.36$\pm$0.20 Mpc away, considerably farther
away than the typically assumed distance of 2.2$\pm$0.6 Mpc. Previously
thought to be an isolated galaxy due to its shorter distance, our new
distance firmly establishes NGC 1569 as a member of the IC 342 group of
galaxies. The higher density environment may help explain the starburst
nature of NGC 1569, since starbursts are often triggered by galaxy
interactions. On the other hand, the longer distance implies that NGC
1569 is an even more extreme starburst galaxy than previously believed. 
Previous estimates of the rate of star formation for stars younger than 
$\la$ 1 Gyr become stronger by more than a factor of 2.  Stars older 
than this were not constrained by previous studies.  The 
dynamical masses of NGC 1569's three super star clusters, which are 
already known as some of the most massive ever discovered, increase by 
$\sim$53\% to 6-7$\times10^5$\msun.
\end{abstract}



\section{Introduction}
Massive starbursts drive the evolution of galaxies at high redshift;
they provide chemical enrichment and thermal and mechanical heating of
both the interstellar and intergalactic medium.  In the past decade, a
large population of star-forming galaxies has been discovered at high
redshift, highlighting the importance of these galaxies on a
cosmological scale.  However, starburst galaxies can only be studied in
detail in the nearby Universe where they are much rarer. The dwarf
irregular galaxy NGC 1569 is one of the closest examples of a true
starburst.  Star formation in NGC 1569 has been studied extensively with
the Hubble Space Telescope (HST; e.g., Greggio et al.~1998, hereafter
G98, Aloisi et al.~2001, Angeretti et al.~2005, hereafter A05), with
results showing that its star formation rates (SFRs) are 2-3 times
higher than in other strong starbursts (e.g.~NGC 1705) and 2-3 orders of
magnitude higher than in Local Group irregulars and the solar
neighborhood, if the SFR per unit area is considered. NGC 1569 is also
home to three of the most massive super star clusters (SSCs) ever
discovered.\looseness=-2

Although it lies on the sky in the same direction as the IC 342 group of
galaxies (see Fig.~2 in Karachentsev et al.~2003), the typically assumed
distance of 2.2$\pm$0.6 Mpc (Israel 1988) is based on the luminosity of
the brightest resolved stars (Ables 1971) and places NGC 1569 well in
front of IC 342 ($D = 3.28 \pm 0.27$ Mpc; Saha, Claver, \& Hoessel
2002). NGC 1569 has therefore generally been viewed as an isolated
starburst galaxy just beyond the outskirts of the Local Group. The
isolated environment makes it more complicated to understand the trigger
of the starburst, since starbursts are often associated with galaxy
interactions and are not generally believed to be internally driven.
However, NGC 1569's distance is relatively uncertain, and a large range
of possible distances exists in the literature.  For example, using
HST/WFPC2 photometry, Makarova \& Karachentsev (2003) attempted to
measure the brightness of stars at the tip of the red giant branch
(TRGB) in NGC 1569 and found, due to the limited number of stars
available at their detection limit, two possible solutions; a short
distance of $1.95\pm0.2$ Mpc or a long distance of $2.8\pm0.2$ Mpc, with
no preference for one over the other.  We note also that distance
determinations based on the brightest resolved stars in dwarf galaxies
are prone to considerable uncertainty due to stochastic effects
(e.g.~Greggio 1986), and that, at the distance of NGC 1569, ground-based
observations suffer from the possibility of misinterpreting star
clusters or pairs of stars as individual stars.

In an effort to better determine its star formation history (SFH), we
have used the HST ACS/WFC to obtain deep $V$- and $I$-band photometry of
NGC 1569. These data also provide for the first time an accurate
distance based on unequivocal identification of the TRGB, which we
present and discuss in this Letter.

\section{Observations and Photometry}
\begin{figure}
\includegraphics[width=0.45\textwidth]{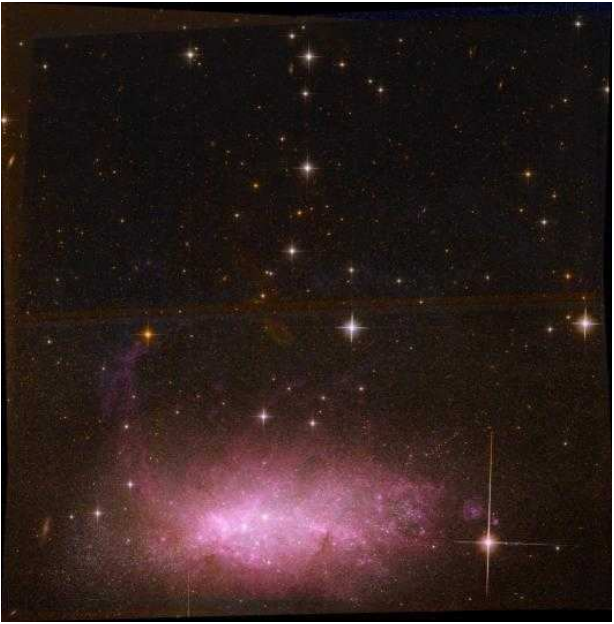}
\caption{ACS/WFC 3-color image of NGC 1569, where the H$\alpha$-, $I$-
and $V$-bands have been colored red, orange, and blue, respectively.
(The faint ellipse near the center of the field is the ghost image of a
bright star).}\label{ngc1569}
\end{figure}

NGC 1569 was imaged in November 2006 and January 2007 with the ACS/WFC
using the F606W ($V$) and F814W ($I$) broad-band filters and the F658N
(H$\alpha$) narrow-band filter ({\it HST} GO-10885, PI: Aloisi). The
H$\alpha$ image was not used for photometry and is thus excluded from
further discussion herein. The total exposure time in the $V$-band was
61716s, comprised of 54 individual images that were taken at two
different orientations.  Similarly, for the $I$-band we have 22 images
with a total exposure time of 24088s, also taken at two orientations.
Images in both filters were dithered using standard patterns to fill in
the gap between the two ACS CCDs and to improve the sampling of the
point spread function (PSF).  For each filter, the individual images
were processed with the most up-to-date version of the ACS pipeline
(CALACS) and then combined into a single image using the MULTIDRIZZLE
software package (Koekemoer et al.~2002).  During the combination, we
fine-tuned the image alignment and corrected for small shifts, rotations
and scale variations between the images. The MULTIDRIZZLE software also
removes cosmic rays and bad pixels and corrects the images for geometric
distortions.  Our final $V$ and $I$ images were resampled to a pixel
size of $0.035\arcsec$ (0.7 times the original ACS/WFC pixel scale) and
cover roughly $3.5\arcmin \times 3.5\arcmin$.  Figure \ref{ngc1569}
shows our 3-color image of NGC 1569.

We performed photometry on the processed images using the stand alone
version of DAOPHOT (Stetson 1987) in the following manner.  To create
the PSF model, in each image we chose $\sim$400 bright stars with good
spatial coverage across the entire image, but avoided the crowded inner
regions of the galaxy.  Nearby neighbors were subtracted from around the
PSF stars and the `cleaned' stars were used to make the final PSF. We
performed PSF fitting on our $V$ and $I$ images using ALLSTAR.  To find
and photometer faint neighbors, we performed a single iteration of
fitting the `known' stars, subtracting them from the frame, searching
for faint companions, and then re-performing the PSF fitting on the
entire original frame using the updated catalog.  The $V$ and $I$
photometry lists were then matched, with a required matching radius of
0.5 pixels, resulting in a photometric catalog of $\sim$445,000 stars.
We corrected for CCD charge transfer efficiency losses following the
prescription in Riess and Mack (2004). The photometry was converted to
the Johnson-Cousins system by following the procedure outlined in
Sirianni et al.~(2005), but using updated ACS zeropoints (Bohlin
2007).\looseness=-2

\section{TRGB Distance to NGC 1569}
\begin{figure}
\includegraphics[width=0.5\textwidth]{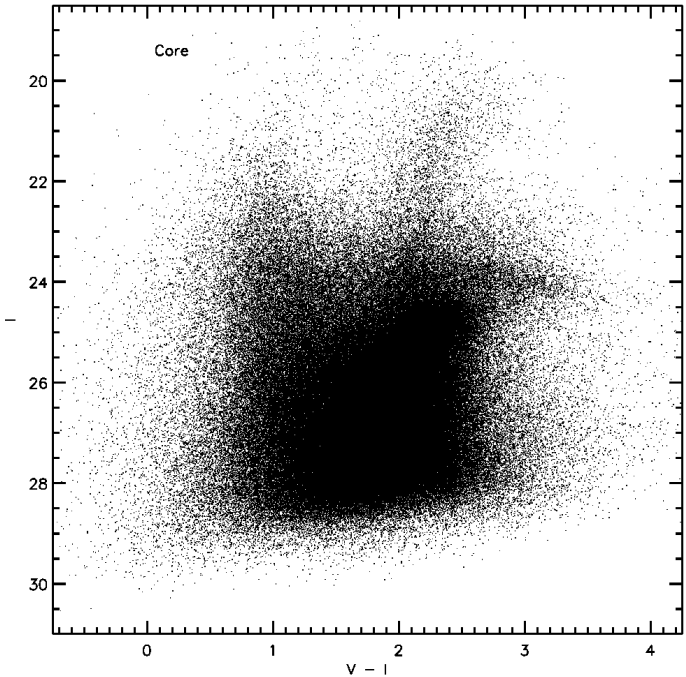}
\caption{CMD composed of all stars in the core of NGC 1569.}
\label{cmd}
\end{figure}

As has been shown by previous authors (e.g.~G98, A05), NGC 1569 has
undergone recent, massive bursts of star formation, and this is
evident in Fig.~\ref{cmd} where we plot the CMD for all stars in the
`core' of NGC 1569.  For the purposes of this paper, we refer to the
bottom half of Fig.~\ref{ngc1569} as the core of NGC 1569 and the top
half as the halo.  Figure \ref{cmd} is dominated by features
associated with young stars; above $I \sim 23.5$, the blue plume,
which samples both the young ($\la 10$ Myr) main sequence as well as
massive ($\ga 9$ \msun) stars on the blue part of the core
helium-burning phase, is clearly visible at $0.6 \la (V-I) \la 1.3$.
In addition, the red plume of supergiants at $1.8 \la (V-I) \la 2.7$
and the blue loop stars (residing between the blue and red plumes) are
indicative of evolved stars with masses between $\sim$ 5 \msun~and 9
\msun. We note also the presence of intermediate-age carbon stars on
the thermally pulsing asymptotic giant branch at $I \sim 24$ and $
(V-I) \ga 2.7$.

\begin{figure}
\includegraphics[width=0.5\textwidth]{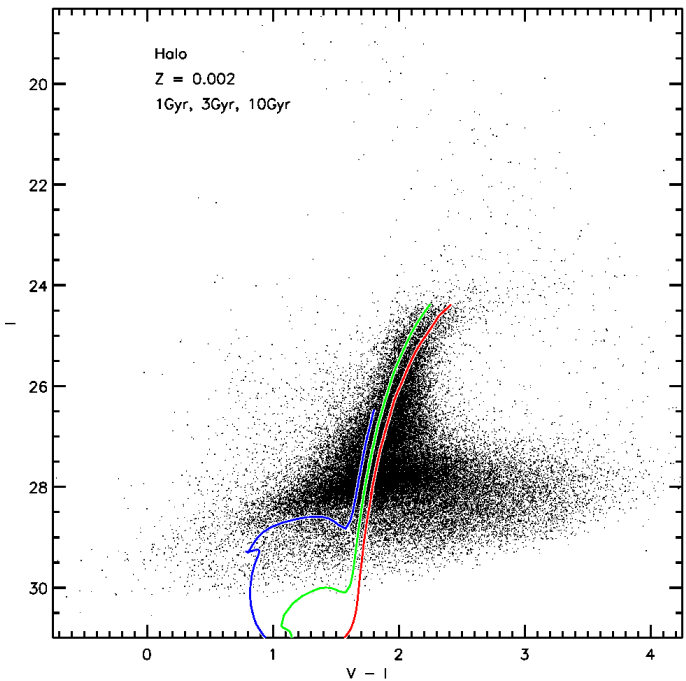}
\caption{CMD of all stars in the halo of NGC 1569.  Overplotted are 
(from left to right) 1 Gyr, 3 Gyr, and 10 Gyr isochrones from Marigo et 
al.~(2008), all with Z = 0.002 ([Fe/H]$\sim$-1).  For clarity, we have 
only plotted evolutionary phases in the isochrones up to the TRGB.}
\label{rgb}
\end{figure}

The recent star formation burst in NGC 1569 has occurred almost
exclusively in the core of the galaxy, as is evidenced by
Fig.~\ref{rgb}, where we have plotted only those stars in the halo of
NGC 1569. In this figure the only obvious feature is the upper 4
magnitudes of the galaxy's RGB, the result of an underlying population
in NGC 1569 with an age $\ga$ 2 Gyr. This feature is visible in the
core CMD as well, but there it is more heavily blended with younger
evolutionary features.

\begin{figure}
\includegraphics[width=0.5\textwidth]{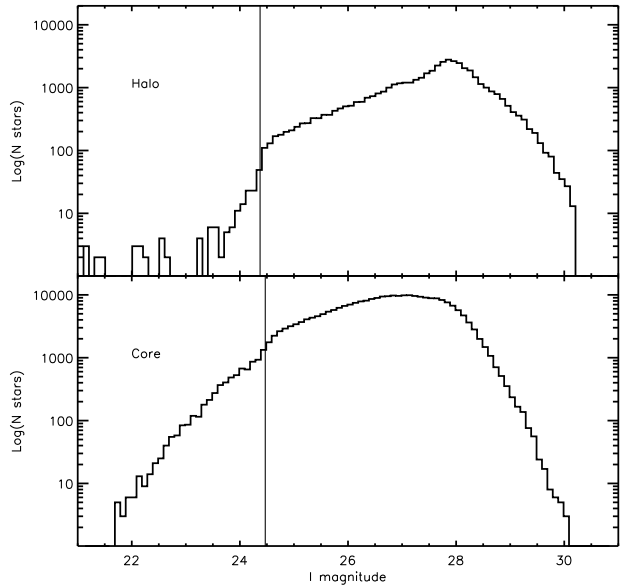}
\caption{LF of NGC 1569's halo ({\it top}) and core ({\it bottom}), with 
the position of the TRGB marked by the solid vertical line.  For the 
core, we have isolated the RGB by only including stars with $6.9-0.2I < 
(V-I) < 2.6$.}
\label{lf}
\end{figure}

The location of the TRGB acts as a discontinuity in the luminosity
function (LF) of a stellar population. We measure the brightness of the
TRGB by following the software developed by one of us (RvdM) and
detailed in Cioni et al.~(2000). The software calculates the first and
second derivatives of the LF, fits Gaussians to their peaks, and then
corrects the peak magnitudes for the small biases induced by LF
smoothing due to photometric errors and binning. In the top panel of
Fig.~\ref{lf} we plot the LF for RGB stars in NGC 1569's halo, and
indicate the identified TRGB at \itrgb~= 24.37$\pm$0.03 (random error). 
The bottom panel shows the core LF and its identified TRGB at \itrgb~=
24.47$\pm$0.01 (in this case we applied a loose CMD color cut to help
isolate RBG stars, although this is not required for our method to
provide valid results). The fainter TRGB magnitude in the core may be
explained by an internal reddening of \ebv~$\sim$ 0.07, or by the
presence of younger and/or more metal-rich RGB stars in the core of the
galaxy. Our TRGB measurements cannot determine the exact cause of this
variation.

A variety of authors have calculated both the foreground reddening
toward and intrinsic reddening in NGC 1569.  Maps of the extinction
due to the Milky Way (MW) provided by Burstein \& Heiles (1982) and
Schlegel et al.~(1998) give values of \ebv~= 0.50 and 0.68 mag,
respectively.  The most commonly adopted extinction for NGC 1569 is
\ebv~= 0.56 (Israel 1988), calculated using integrated UV photometry.
We note that we adopt the reddening law of Cardelli et al.~(1989),
where \av~= 3.1\ebv, and \ai~= 0.479\av.  We test the range of
reddening values by comparing our photometry to the Padova isochrones
(Marigo et al.~2008), focusing on matching the slope and color of the
observed and theoretical RGBs in the halo. The {\it slope} of the
observed CMD is best matched by isochrones with Z = 0.002 ($\sim$
0.1\zsol).  Isochrones that are more metal-poor are too vertical to
match the RGB, while more metal-rich isochrones have too much
curvature. The {\it color} of the data best matches the Z = 0.002
isochrones if we use the lowest reddening value, \ebv~= 0.50.  In
Fig.~\ref{rgb} we illustrate this agreement by overplotting isochrones
for stellar populations with ages of 1, 3, and 10 Gyr, Z = 0.002
([Fe/H] $\sim -1$), and \ebv~= 0.50.  Adopting a reddening value on
the low end of the published values is reasonable for the following
reasons.  As discussed by Rela\~{n}o et al.~(2006), the Schlegel et
al.~(1998) foreground reddening is higher than the {\it total}
extinction derived by most authors (e.g.~Israel 1988, Origlia et
al.~2001) and is likely an overestimate of the reddening at the
position of NGC 1569.  In addition, since the value calculated by
Israel (1988) is based on integrated UV photometry of the core of NGC
1569, it is probably a measure of both the foreground reddening and
extinction due to the gas and dust associated with the recent burst of
star formation.  Since the recent star formation in NGC 1569 is
concentrated in the core, extinction in the halo should be lower.\looseness=-2

Using synthetic CMDs, Barker, Sarajedini, and Harris (2004) studied
the reliability of the TRGB as a distance indicator for stellar
populations with complex star formation histories. They found that the
$I$-band TRGB absolute magnitude \mitrgb~= $-4.0 \pm$ 0.1 independent
of SFH, provided that the combination of age, metallicity, and
multiple bursts are such that the median {\it dereddened} color of the
RGB stars $\sim$ 0.5 mag below the TRGB is $(V-I)_0 \la 1.9$. For NGC
1569, we find the median $(V-I)_0$ = 1.4 for all RGB stars between 0.4
and 0.6 mag below the tip. It is therefore appropriate to assume
\mitrgb~= $-4.0 \pm$ 0.1 for our distance calculations. Combined with
\itrgb~= 24.37$\pm$0.03 and \ebv~= 0.50$\pm$0.05 in the halo this
yields $(m-M)_0 = 27.63\pm0.13$, or $D = 3.36 \pm 0.20$ Mpc.

\section{Implications of a Longer Distance}

The IC 342 group of galaxies has a mean distance of 3.35$\pm$0.09 Mpc
with a line-of-sight depth of 0.25 Mpc (1$\sigma$; Karachentsev 2005,
hereafter K05). Our new HST TRGB distance to NGC 1569 places it at the
same distance as the group, and only slightly farther away than IC 342
itself (3.28$\pm$0.27 Mpc; Saha et al.~2002). Given the apparent
separation of 5$\fdg$45, and applying the law of cosines, we find that
NGC 1569 is the 4th closest galaxy to IC 342, with a physical
separation of $326_{-25}^{+262}$ kpc. In the rest frame of the Local
Group, the IC 342 group has $\langle V_{LG} \rangle$ = 226$\pm$18 km
s$^{-1}$ (standard error), with a dispersion of 54 km s$^{-1}$ (K05).
NGC 1569's radial velocity, $V_{LG}$ = 88 km s$^{-1}$, is 2.5$\sigma$
from the group average. While NGC 1569 may simply be a statistical
outlier in the velocity distribution, an alternative explanation is
that NGC 1569 may be falling into the IC 342 group for the first
time. Either way, NGC 1569 is not an isolated galaxy, so its starburst
may be due to tidal interactions with other galaxies in the IC 342
group.

The increased distance of NGC 1569 alters the existing estimates for
its SFR and SSC masses that were based on the old distance of $2.2$
Mpc. SSCs are young, massive clusters that are bright enough to be
easily observed in detail out to beyond the Local Group.  Their
masses, which are several orders of magnitude higher than typical
Galactic open clusters, are large enough that SSCs are likely to
survive for the entire lifetime of their host galaxy. SSCs are thus
possible precursors to globular clusters. The Milky Way and Large
Magellanic Cloud (LMC) host massive young clusters up to $\sim
10^5$\msun. However, the starburst in NGC 1569 has spawned three SSCs
that are even more massive: NGC 1569-A1, NGC 1569-A2 (hereafter A1 and
A2), and NGC 1569-B.

The most recent mass estimate for any of these clusters was calculated
for NGC 1569-B by Larsen et al.~(2008).  Combining HST photometry with
high-resolution spectroscopy, and assuming that NGC 1569-B is in
virial equilibrium, they find $M_{vir} = (4.4 \pm 1.1) \times
10^5$\msun for $D=2.2$ Mpc. The clusters A1 and A2 were originally
thought to be a single cluster, before being resolved into two
components (de Marchi et al.~1997). However, mass estimates still
typically treat A1 and A2 as a single cluster (e.g.~Anders et
al.~2004).  Gilbert and Graham (2002) provided the most recent
calculation of masses for A1 and A2 individually.  They obtained a
high-resolution spectrum that samples both A1 and A2 simultaneously
and found that the cross-correlation function between their spectrum
and standard stars is best fit by two components having similar
line-of-sight velocity dispersions, but different peak velocities.
The resulting masses of the two components are 3.9 $\times 10^5$ and
4.4 $\times 10^5$ \msun for $D=2.2$ Mpc (their method does not
determine which component has which mass). Given the linear
relationship between dynamical mass and distance, our longer distance
to NGC 1569 increases all these mass estimates by 53\%.  Thus, the
mass of NGC 1569-B becomes $6.7 \times 10^5$\msun, and A1 and A2 are
$6.0$ and $6.7 \times 10^5$\msun.

Regarding the star formation in NGC 1569, the most recent work was by
A05. Using HST NICMOS photometry they identified three major epochs of
star formation, 1.3$-$3.7 $\times 10^7$ yr ago, 4$-$15 $\times 10^7$ yr
ago, and $\sim$ 1 Gyr ago, with SFRs of $\sim$ 3.2, $\sim$ 1.1, and
$\sim$ 0.8 \msun yr$^{-1}$ kpc $^{-2}$, respectively. The primary SFR
calculations from A05 were based on an assumed distance of 2.2 Mpc, but
they also calculated SFRs using a distance of 2.9 Mpc.  By adopting the
longer distance, SFRs increase by a factor of 2, and the epochs of star
formation shift to younger ages (e.g., 0.8$-$2.7 $\times 10^7$ yr ago
for the youngest epoch).  Similarly, G98 found that a shift in the
distance of NGC 1569 to 4 Mpc from 2.2 Mpc increased the recent SFR by a
factor of 2.8 and suggested that the youngest epoch of star formation
ended more recently (5 Myr ago as compared to 10 Myr ago).  Note that
these results only apply to stars younger than $\sim$1 Gyr as older
stars were not constrained by these studies.  The present-day SFR, based
on H$\alpha$ observations by Hunter \& Elmegreen (2004), increases from
0.32 \msun yr$^{-1}$ to 0.75 \msun yr$^{-1}$ due to their use of 2.5 Mpc
as the distance to NGC 1569.  A05 did not need to include a significant
epoch of star formation older than $\sim$ 1 Gyr in their models due to a
deficiency of RGB stars detectable in their data.  In contrast, our
data, which are significantly deeper and cover a much larger area, show
a well populated, fully formed RGB, suggesting that star formation in
NGC 1569 began $\ga$ 2 Gyr ago.  In a future paper we will make use of
the depth and areal coverage of our photometry to re-derive the SFRs
with a technique similar to A05 and characterize the variations in the
SFH of NGC 1569 as a function of both time and location.

\acknowledgments
Support for proposal GO-10885 was provided by NASA through a grant from 
STScI, which is operated by AURA, Inc., under NASA contract NAS 5-26555.\\
%% copy editing.  Individual instruments can be provided in parentheses,
%% after the keyword, but they will not be verified.
%
Facilities: \facility{HST (ACS)}.

%\appendix

%\section{Appendix material}


\begin{thebibliography}{}
\bibitem[]{} Ables, H.D.  1971, Publications of the U.S.~Naval 
Observatory Second Series, 20, 126
\bibitem[]{} Aloisi, A., et al.  2001, \aj, 121, 1425
%Clampin, M., Diolaiti, E., Greggio, L., Leitherer, C., Nota, A., 
%Origlia, L., Parmeggiani, G., \& Tosi, M.  2001, \aj, 121, 1425
\bibitem[]{} Anders, P., de Grijs, R., Fritze-v.Alvensleben, U., \& 
Bissantz, N.  2004, \mnras, 347, 17
\bibitem[]{} Angeretti, L., Tosi, M., Greggio, L., Sabbi, E., Aloisi, 
A., \& Leitherer, C.  2005, \aj, 129, 2203 (A05)
\bibitem[]{} Barker, M.K., Sarajedini, A., \& Harris, J.  2004, \apj, 
606, 869
\bibitem[]{} Bohlin, R.C.  2007 in ASP Conf. Ser. 364, The Future of 
Photometric, Spectrophotometric, and Polarimetric Standardization, ed. 
C. Sterken (San Francisco: ASP), 315
\bibitem[]{} Burstein, D. \& Heiles C.  1982, \aj, 87, 1165 
\bibitem[]{} Cardelli, J.A., Clayton, G.C., \& Mathis, J.S.  1989, \apj, 
345, 245
\bibitem[]{} Cioni, M.-R. L., van der Marel, R.P., Loup, C., \& Habing, 
H.J.  2000, \aap, 359, 601
%\bibitem[]{} Clark, J.S., Negueruela, I., Crowther, P.A., \& Goodwin, 
%S.P.  2005, \aap, 434, 949
\bibitem[]{} de Marchi, G., Clampin, M., Greggio, L., Leitherer, C., 
Nota, A., \& Tosi, M.  1997, \apjl, 479, 27
\bibitem[]{} Gilbert, A.M., \& Graham, J.R.  2002, in IAU Symp. 207, 
Extragalactic Star Clusters, ed. D. Geisler, E.K. Grebel, \& D. Minniti 
(San Francisco: ASP), 471
%\bibitem[]{} Girardi, L., Bressan, A., Bertelli, G., \& Chiosi, C. 
%2000, \aaps, 141, 371
\bibitem[]{} Greggio, L. 1986, \aap, 190, 111
\bibitem[]{} Greggio, L., Tosi, M., Clampin, M., De Marchi, G., 
Leitherer, C., Nota, A. \& Sirianni, M.  1998, \apj, 504, 725 (G98)
\bibitem[]{} Hunter, D.A. \& Elmegreen, B.G.  2004, \aj, 128, 2170
\bibitem[]{} Israel, F.P.  1988, \aap, 194, 24
\bibitem[]{} Karachentsev, I.D., Sharina, M.E., Dolphin, A.E., \& 
Grebel, E.K.  2003, \aap, 408, 111
\bibitem[]{} Karachentsev, I.D.  2005, \aj, 129, 178
\bibitem[]{} Koekemoer, A. M., Fruchter, A. S., Hook, R. N., \& Hack, W. 
J. 2003, HST Calibration Workshop (Baltimore: STScI), 337
\bibitem[]{} Larsen, S.S., Origlia, L., Brodie, J., \& Gallagher, J.S. 
III  2008, \mnras, 383, 263
\bibitem[]{} Marigo, P., Girardi, L., Bressan, A., Groenewegen, 
M.~A.~T., Silva, L., \& Granato, G.~L., \aap, 482, 883
\bibitem[]{} Makarova, L.N. \& Karachentsev, I.D.  2003, Astrophysics, 
46, 144
% \bibitem[]{} Mengel, S. \& Tacconi-Garman, L.E.  2008, astro-ph/0803.4471
\bibitem[]{} Origlia, L., Leitherer, C., Aloisi, A., Greggio, L., \& 
Tosi, M. 2001, \apj, 122, 815
\bibitem[]{} Riess, A. \& Mack, J.  2004, ISR ACS 2004-006
\bibitem[]{} Rela\~{n}o, M., Lisenfeld, U., Vilchez, J. M., \& Battaner, 
E.  2006, \aap, 452, 413
\bibitem[]{} Saha, A., Claver, J., \& Hoessel, J.G. 2002, \aj, 124, 839
\bibitem[]{} Schlegel, D.J., Finkbeiner, D.P., Davis, M.  1998, \apj, 
500, 525
\bibitem[]{} Sirianni, M. et al. 2005, \pasp, 117, 1049
\bibitem[]{} Stetson, P.B.  1987, \pasp, 99, 191S
%\bibitem[]{} van den Bergh, S.  1999, \pasp, 111, 1248
\end{thebibliography}
\end{document}